\documentclass{PoS}

\usepackage{amsmath}
\usepackage{amssymb}
\usepackage{array}
\usepackage{calc}
\usepackage{longtable}
\usepackage{multirow,booktabs}
\usepackage{relsize}
\usepackage{pstricks}
\usepackage{graphicx}
\usepackage{xspace}
\usepackage{units}
\newcommand{\MCatNLO}{M\protect\scalebox{0.8}{C}@N\protect\scalebox{0.8}{LO}\xspace}

\newcommand{\MEPSatLO}{M\protect\scalebox{0.8}{E}P\protect\scalebox{0.8}{S}@L\protect\scalebox{0.8}{O}\xspace}
\newcommand{\MEPSatNLO}{M\protect\scalebox{0.8}{E}P\protect\scalebox{0.8}{S}@N\protect\scalebox{0.8}{LO}\xspace}

\newcommand{\Qcut}{\ensuremath{Q_\mathrm{cut}}}

\newcommand{\Sherpa}{S\protect\scalebox{0.8}{HERPA}\xspace}

\newcommand\ATLAS{\atlas}
\newcommand\atlas{A\protect\scalebox{0.8}{TLAS}\xspace}
\newcommand\CMS{\cms}
\newcommand\cms{C\protect\scalebox{0.8}{MS}\xspace}
\title{Bottom-quark mass effects in associated production with $Z$ and $H$ bosons}

\ShortTitle{Bottom-quark mass effects in associated production with $Z$ and $H$ bosons}

\author{\speaker{Davide Napoletano}\\
  Institute of Particle Physics Phenomenology, Durham University\\
  E-mail: \email{davide.napoletano@durham.ac.uk}}
\author{Frank Krauss\\
  Institute of Particle Physics Phenomenology, Durham University\\
  E-mail: \email{frank.krauss@durham.ac.uk}}
\author{Steffen Schumann\\
  II. Physikalisches Institut, Georg-August-Universit\"at G\"ottingen, 37077 G\"ottingen, Germany \\
  E-mail: \email{steffen.schumann@phys.uni-goettingen.de}}

\abstract{In this study, predictions obtained in the four and in the five
  flavour schemes are compared for two important processes involving heavy
  flavours at the LHC: the production of a $Z$ or a Higgs boson in association
  with $b$ quarks. In particular we obtain predictions with \Sherpa's
  \MCatNLO implementation for the four--flavour scheme, treating the $b$'s
  as massive, and with multijet merging at leading and next-to leading
  order for the five--flavour scheme.

  While differences between the two schemes, at the inclusive level,
  are well understood from resummation of possibly large logs into
  the $b$-PDFs, differences in shape present a major problem
  for experimental measurements. We make use of data for $Z+b(\bar{b})$
  production at the $7$ TeV LHC to exhibit strengths and weaknesses of the
  different approaches and we use these results to validate predictions for
  $b$-associated Higgs-boson production at the 13 TeV Run II.}

\FullConference{XXV International Workshop on Deep-Inelastic Scattering
  and Related Subjects\\
  3-7 April 2017\\
  University of Birmingham, UK}

\begin{document}

\section{4F or 5F scheme?}
The study of processes that involve heavy quarks in the LHC environment
is of great importance for theorists and experimentalists alike.
From the theory view point, processes like $Z/W b$ are sensitive
to the heavy-flavour content of the proton and are thus used to determine
the $b$-PDF, which in turn is necessary to make predictions for processes
like $b\bar{b}\rightarrow H$. In addition, Higgs production in bottom-quark
fusion, although being characterised by a very small cross section in the
SM, can be sensitive for BSM scenarios where the bottom Yukawa coupling
is enhanced.

On the other hand, experimentalists, under the name
of {\it theory uncertainties}, take the differences between a scheme
in which the $b$ quark is treated as a massive, decoupled particle,
and one in which it is treated on the same footing as
any other light quark, as input. These two approaches
are usually called 4F and 5F scheme respectively.

Although in principle, the 4F and the 5F schemes should have
only subleading differences, historically they have been found
to largely disagree both at the level of total inclusive cross section and
at the level of differential distributions~\cite{Maltoni:2012pa,Harlander:2003ai,Aad:2014dvb}.
In this study we reconsider these differences. Firstly we
examine the origin of the difference at the inclusive
level in the case of Higgs production
in bottom-quark fusion~\cite{Forte:2015hba,Forte:2016sja}. Secondly we set-up
the two schemes in such a way that they can be compared
likes with likes in differential distributions~\cite{Krauss:2016orf}.

\section{Resummation vs mass effects}
The difference between the two schemes can be expressed in the following,
short, way. The 4FS accounts for mass, power suppressed, terms exactly
at the accuracy of the perturbative order of the calculation. On the other
hand the 5FS, neglects all power corrections proportional to the mass
of the heavy quark and resums logs of $\mu_F/m_b$ to all order
through DGLAP equations. This is achieved by defining a perturbatively generated $b$-PDF.
In practice, this means that the leading-log (LL) term of the 5F scheme
expression corresponds to the LO expression in the 4FS up to power suppressed terms,
and so on for higher orders.

\begin{figure}[htb]
  \centering{
    \includegraphics[width=0.7\textwidth]{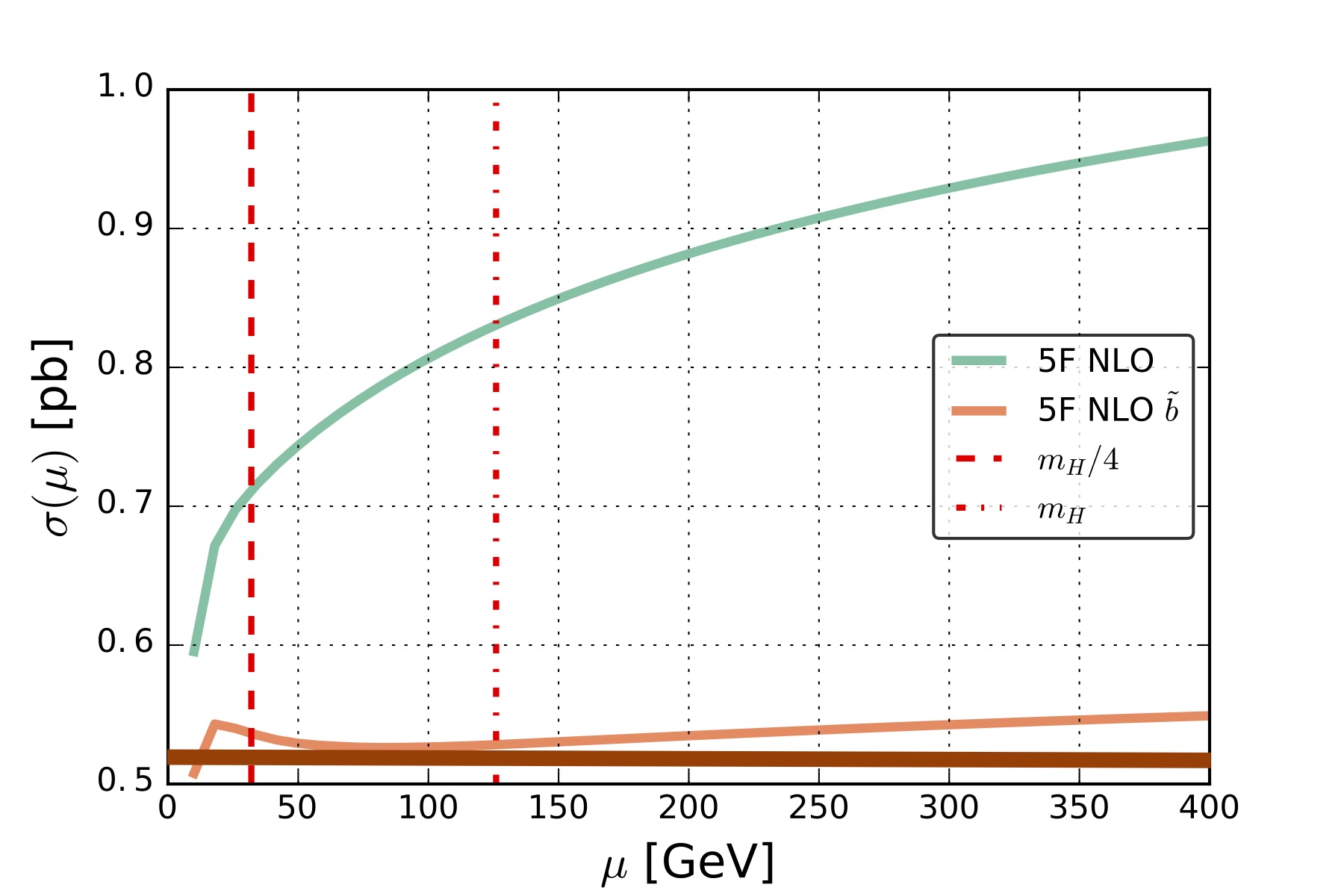}
  }
  \caption{Comparison between a full 5FS and a 5FS in which an expanded $b$-PDF, or $\tilde{b}$,
    is used at NLL. This latter scheme is further compared to the 4FS result (obtained for the
  fixed scale choice $\mu_R=\mu_F=\frac{m_H+2m_b}{4}$) at NLO, brown solid line. }
  \label{fig:btilde}
\end{figure}

The question then reduces to: what is the source of the differences we see, are they mainly due to the
resummation of the logs, or to the exact inclusion of power suppressed terms?
An answer to this question can be obtained by comparing the 5FS prediction
for Higgs production in $b$-quark fusion, with an expanded $b$-PDF to a given order,
and the corresponding 4F prediction. As an example, in Fig.~\ref{fig:btilde},
we show a comparison between the 5F prediction at NLO, using the full $b$-PDF,
the same NLO prediction obtained with a NLL-expanded $b$-PDF, also called $\tilde{b}$,
and the baseline 4FS NLO prediction.

There are two main conclusions that can be taken from Fig.~\ref{fig:btilde}.
The difference between the 5F scheme with an expanded $b$-PDF
and the corresponding 4F scheme calculation is indeed very small,
thus stating that power-corrections are indeed suppressed when looking at total rates,
while the main difference
is generated by the resummation of higher order logs. In addition, choosing a lower
factorisation and renormalisation scale partially reduces this difference.

\begin{figure}[htb]
  \centering{
    \includegraphics[width=0.7\textwidth]{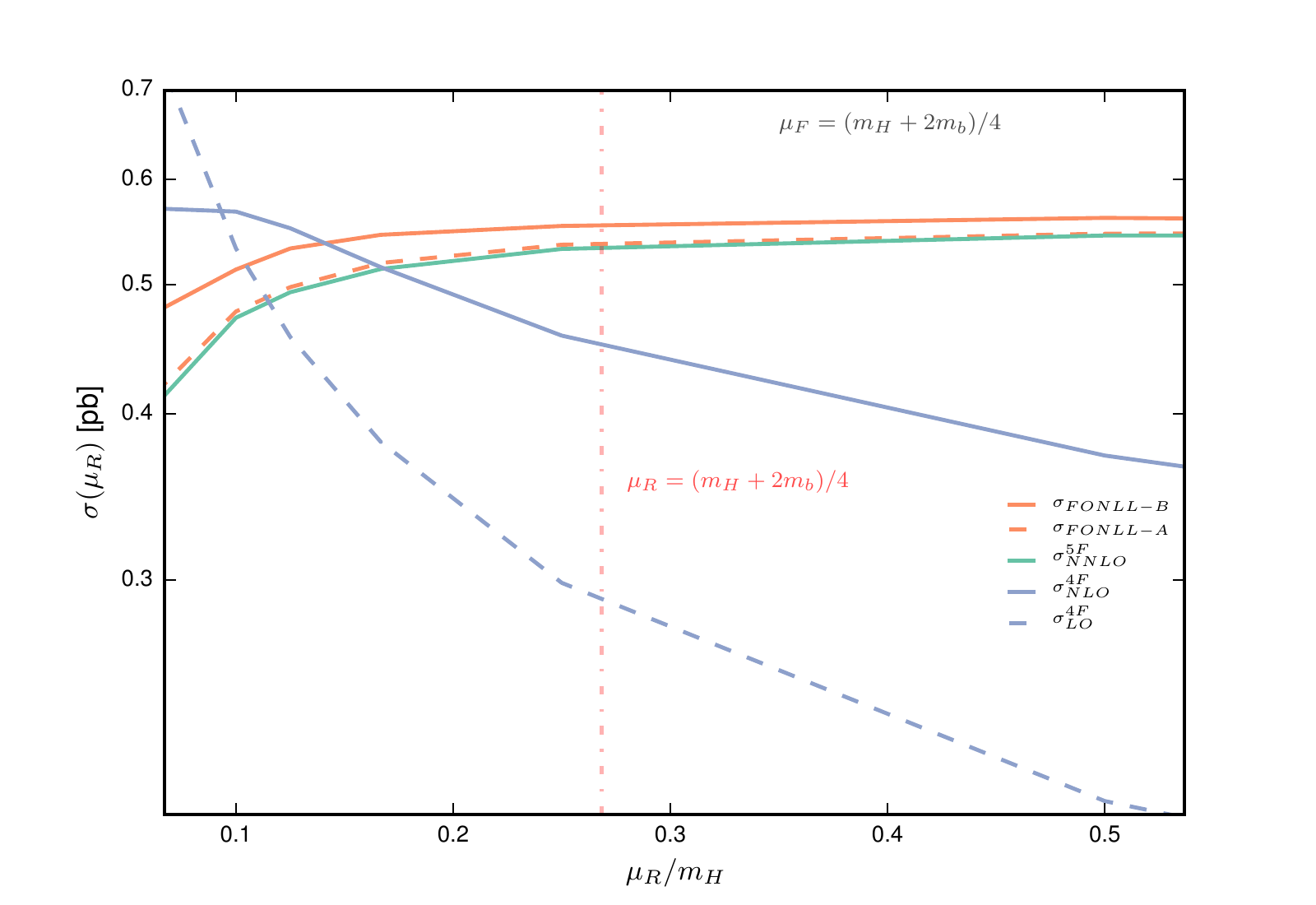}
    \includegraphics[width=0.7\textwidth]{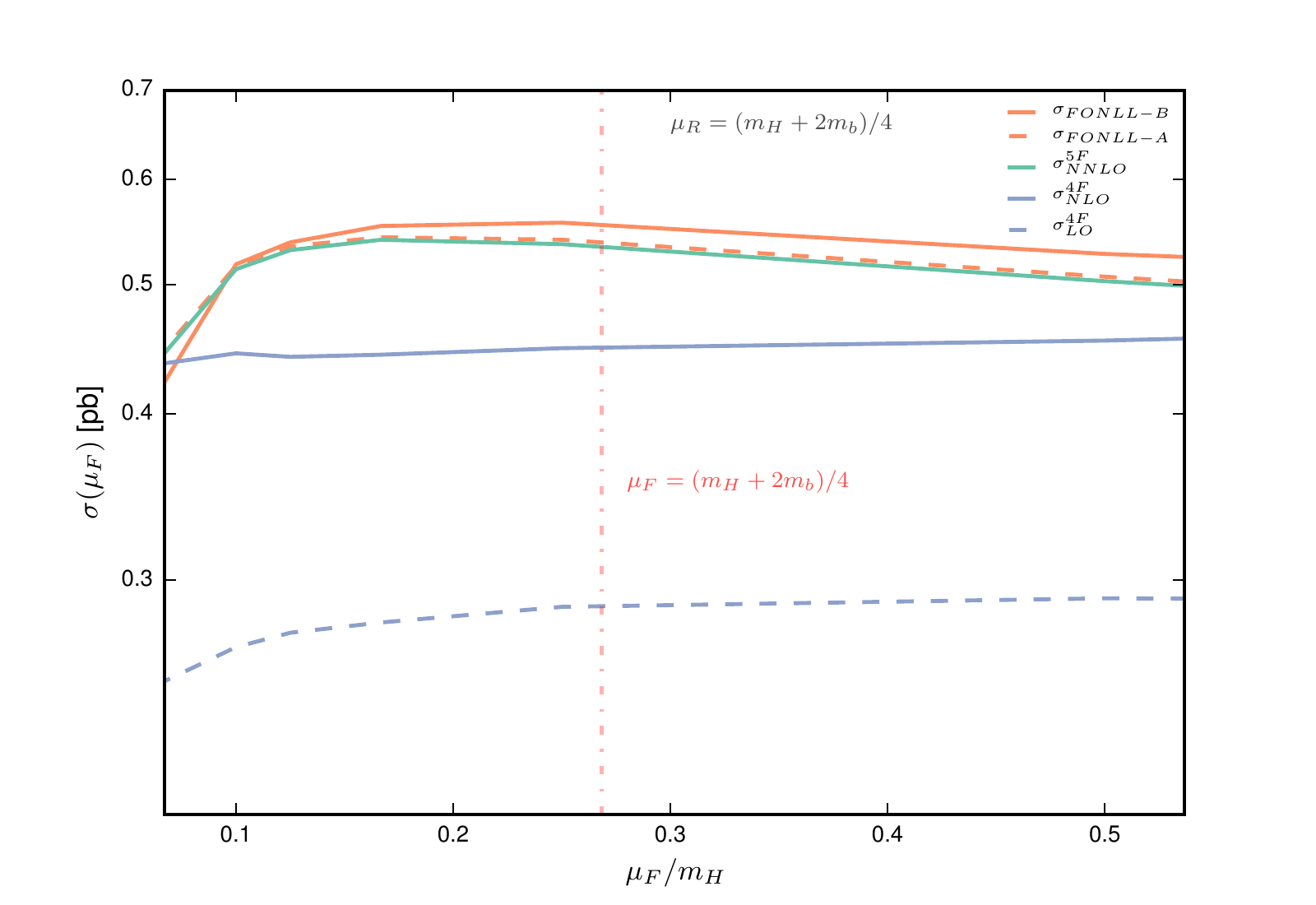}
  }
  \caption{FONLL prediction at LO+NLL (FONLL-A) and at NLO+NNLL (FONLL-B) compared to the 4FS and the 5FS.}
  \label{fig:fonll}
\end{figure}

The conclusion that mass corrections for this process are small, can also be obtained
by matching the 4F and the 5F scheme. This has been done at NLO+NNLL accuracy
using the FONLL~\cite{Forte:2015hba,Forte:2016sja,Cacciari:1998it,Forte:2010ta}
method and a method based on EFT~\cite{Bonvini:2015pxa,Bonvini:2016fgf},
and both methods have been found to agree. Results obtained in the FONLL approach
are shown in Fig.~\ref{fig:fonll}. 

It is thus clear that resummation of large logs is the dominant effect, and not including
them has a larger impact than including mass corrections.

This is however only true
for inclusive observables, like the total inclusive cross section. In order to study
possible effects on differential observables, where a matching has not been
performed yet, we need data. We therefore take as an example the production
of a $Z$ boson in association with at least one or two $b$-jets, and data are
taken from the \ATLAS~\cite{Aad:2014dvb} and \CMS~\cite{Chatrchyan:2013zja} collaborations.

\section{$Zb\bar{b}$ @ 7 TeV}
While a detailed description of the simulation set-up can be found in~\cite{Krauss:2016orf},
we briefly describe here the three samples presented in the following plots.

\begin{description}
\item[4F NLO (4F \MCatNLO):] In the {\em four--flavour scheme}, $b$-quarks
  are consistently treated as {\em massive} particles, only appearing in
  the final state.  As a consequence, $b$-associated $Z$- and $H$-boson
  production proceeds through the parton-level processes $gg\to Z/H+b\bar{b}$,
  and $q\bar{q}\to Z/H+b\bar{b}$ at Born level.  \MCatNLO matching is
  obtained by consistently combining fully differential NLO QCD calculations
  with the parton shower, cf.~\cite{Frixione:2002ik,Hoeche:2011fd}.
\item[5F LO (5F \MEPSatLO):] In the {\em five--flavour scheme} $b$-quarks
  are {\em massless} particles in the {\em hard matrix element}, while they
  are treated as massive particles in both the initial- and final-state
  {\em parton shower}.  
  In the \MEPSatLO~\cite{Hoeche:2009rj} samples we merge $pp \rightarrow H/Z$
  plus up to three jets at leading order; this includes, for instance, the
  parton--level processes $b\bar{b} \to Z/H$, $gb\to Z/H b$,
  $gg\to Z/H b\bar{b}$, $\dots$.  To separate the various matrix-element
  multiplicities, independent of the jet flavour, a jet cut of
  $\Qcut = 10$~GeV is used in the $Z$ case while $\Qcut = 20$~GeV is
  employed in $H$-boson production.
\item[5F NLO (5F \MEPSatNLO):] In the 5FS \MEPSatNLO
  scheme~\cite{Gehrmann:2012yg,Hoeche:2012yf}, we account for quark masses in
  complete analogy to the LO case: the quarks are treated as massless in the
  hard matrix elements, but as massive in the initial- and final-state parton
  showering.  Again, partonic processes of different multiplicity are merged
  similarly to the \MEPSatLO albeit retaining their next-to-leading-order
  accuracy.  In particular, we consider the merging of the processes
  $pp\rightarrow H/Z$ plus up to two jets each calculated with \MCatNLO
  accuracy further merged with $pp\rightarrow H/Z + 3j$ calculated at
  \MEPSatLO.
\end{description}

\begin{figure}[htb]
  \centering{
    \includegraphics[width=0.45\textwidth]{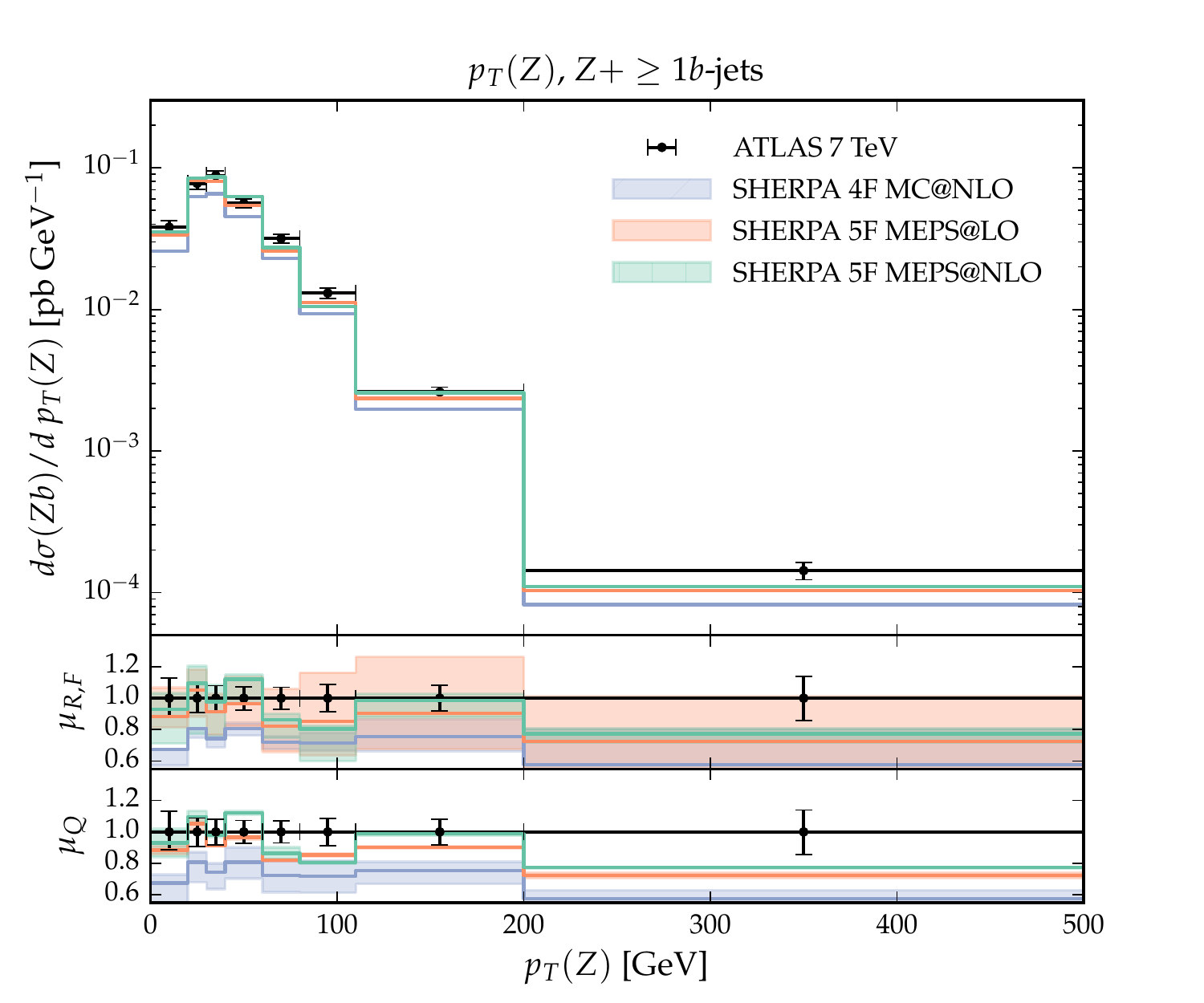}
    \includegraphics[width=0.45\textwidth]{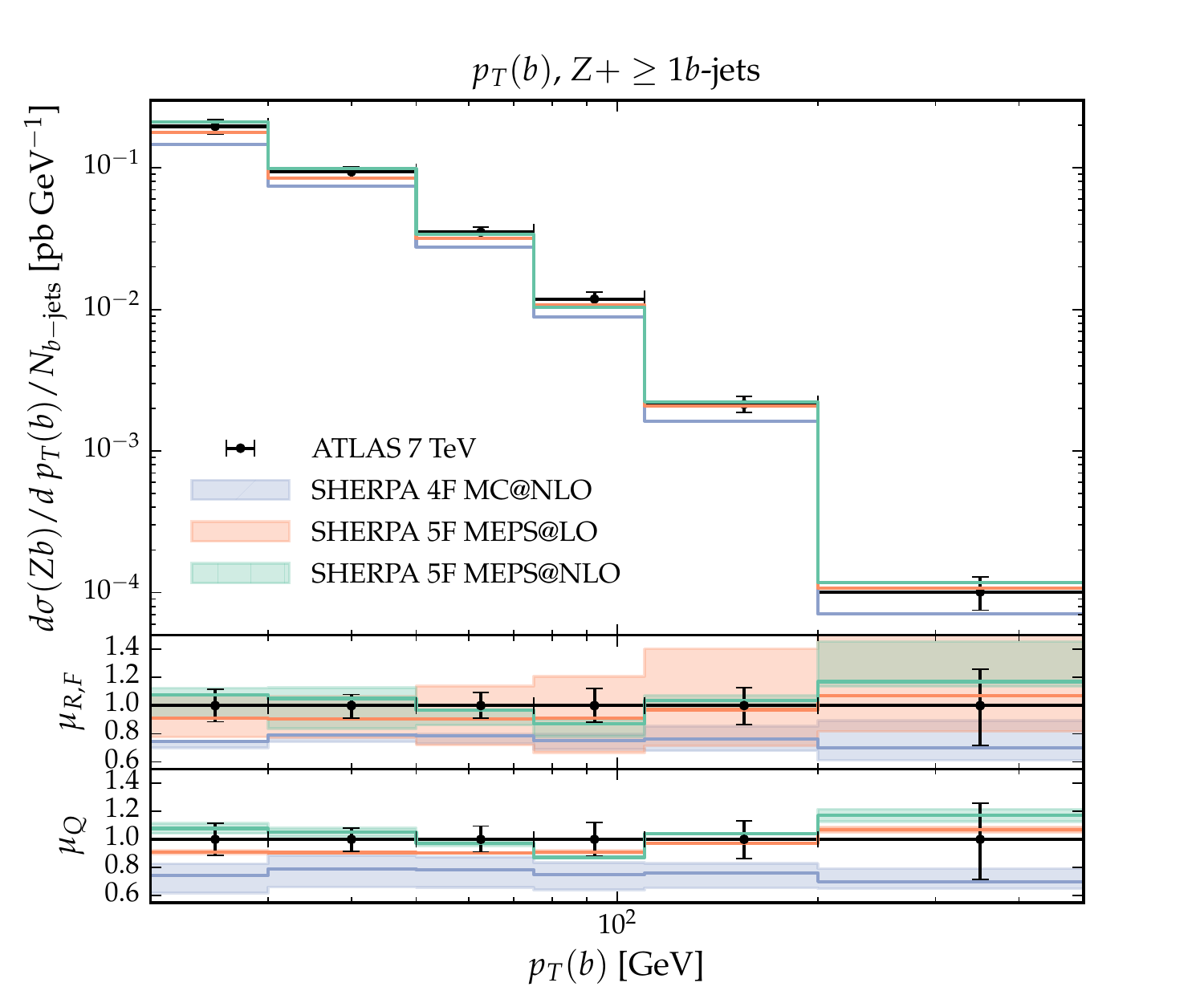}
  }
  \caption{$p_T$ of the $Z$ boson and of the leading $b$-jet in the $\geq 1 b$-jet sample
  for the \ATLAS detector}
  \label{fig:z1b}
\end{figure}

\begin{figure}[htb]
  \centering{
    \includegraphics[width=0.45\textwidth]{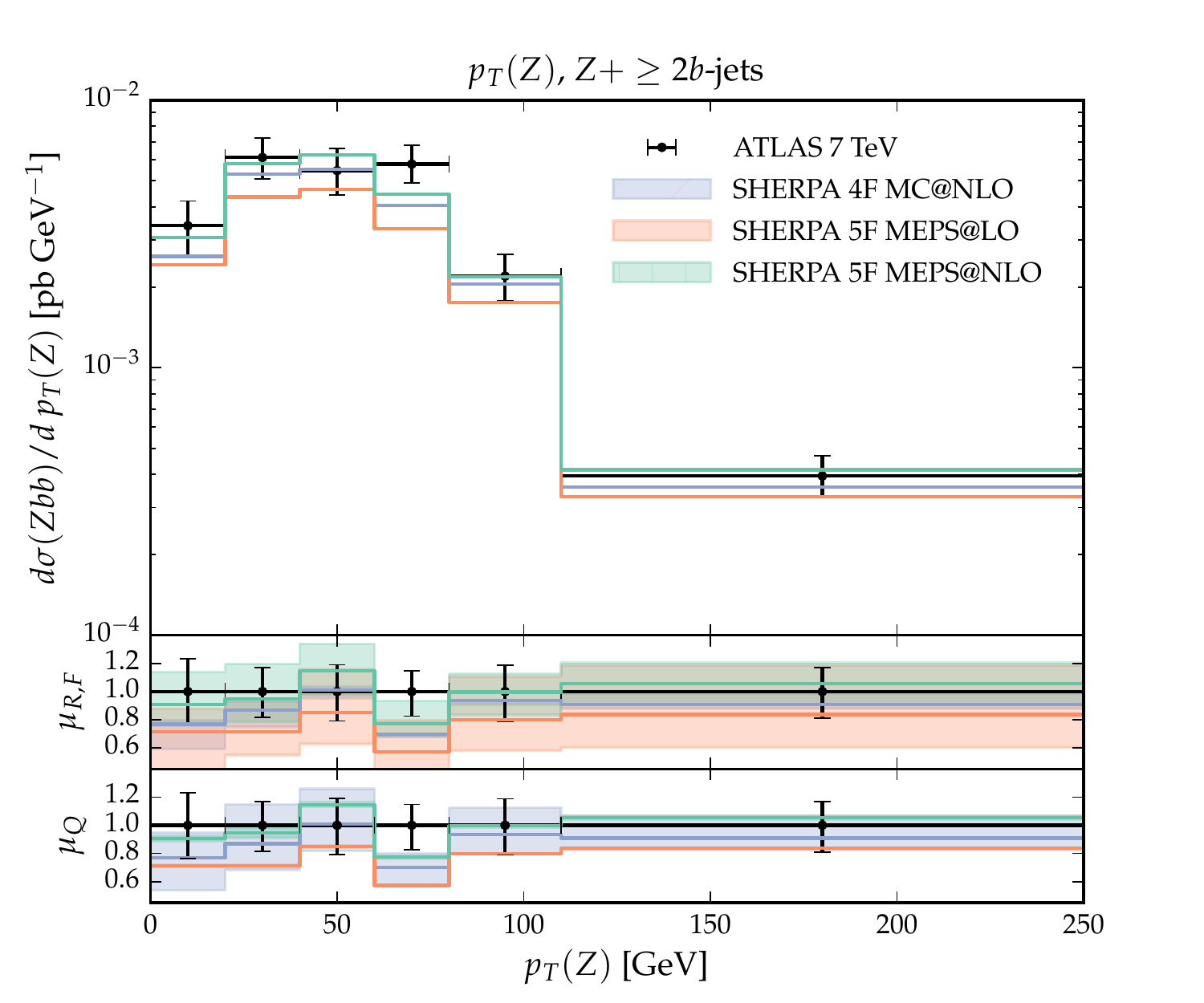}
    \includegraphics[width=0.45\textwidth]{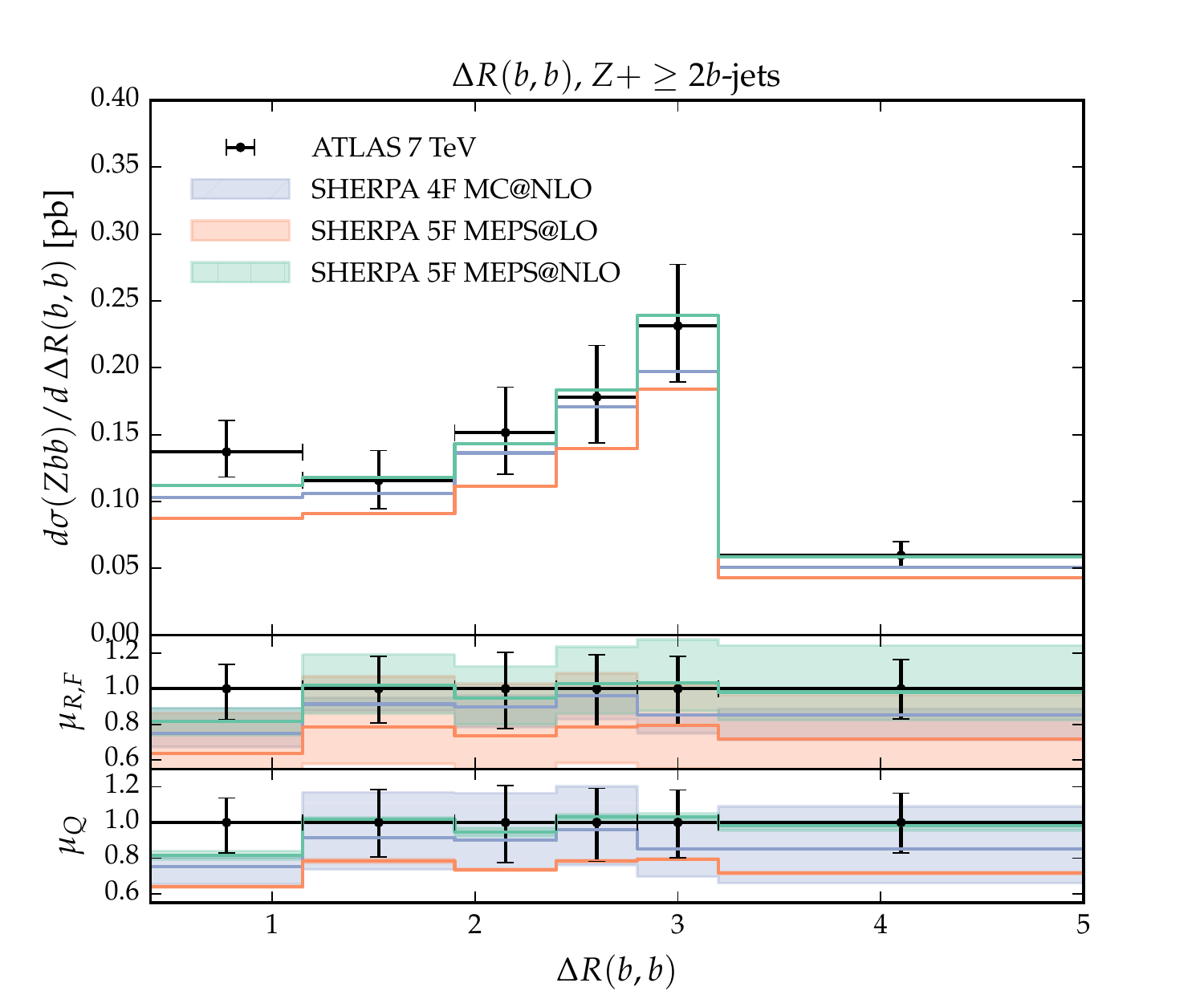}
  }
  \caption{$p_T$ of the $Z$ boson and azimuthal distance between the two leading $b$-jets
    in the $\geq 2 b$-jet sample for the \ATLAS detector}
  \label{fig:z2b}
\end{figure}

Results for the case of the \ATLAS detector are shown in Figs.~\ref{fig:z1b}
for samples that exhibit at least one additional $b$-jet, and in Figs.~\ref{fig:z2b}
for samples with at least two $b$-jets tagged in the final state. Results for the \CMS
detector can be found in~\cite{Krauss:2016orf}, and yield similar conclusions.

For both samples, we find that the 5FS \MEPSatNLO prediction, the one that has the
resummation of the initial state logs, is the one that performs best, in both
normalisation and shape. The 4FS and the 5FS \MEPSatLO show good agreement
in normalisation in the $\geq 2 b$-jets case and in the $\geq 1 b$-jet case, respectively,
while in the opposite cases they both undershoot data, by a largely flat $\sim 20\%$.
All in all the three sample perform roughly at the same level in terms
of shapes, yielding essentially flat $K$-factors to one another.
The good agreement is essentially due to the inclusion of the necessary
higher multiplicity matrix element corrections in the 5FS for the $\geq 2 b$-jets case,
and with the NLO matching to the shower in the 4FS in the $\geq 1 b$-jet case. It is also
worth noticing that very low $\Delta R(b,b)$ effects are shadowed by the inclusion
of hadronisation effects, which are necessary to compare with the available data points.

\section{$Hb\bar{b}$ @ 13 TeV}
\begin{figure}[htb]
  \centering{
    \includegraphics[width=0.45\textwidth]{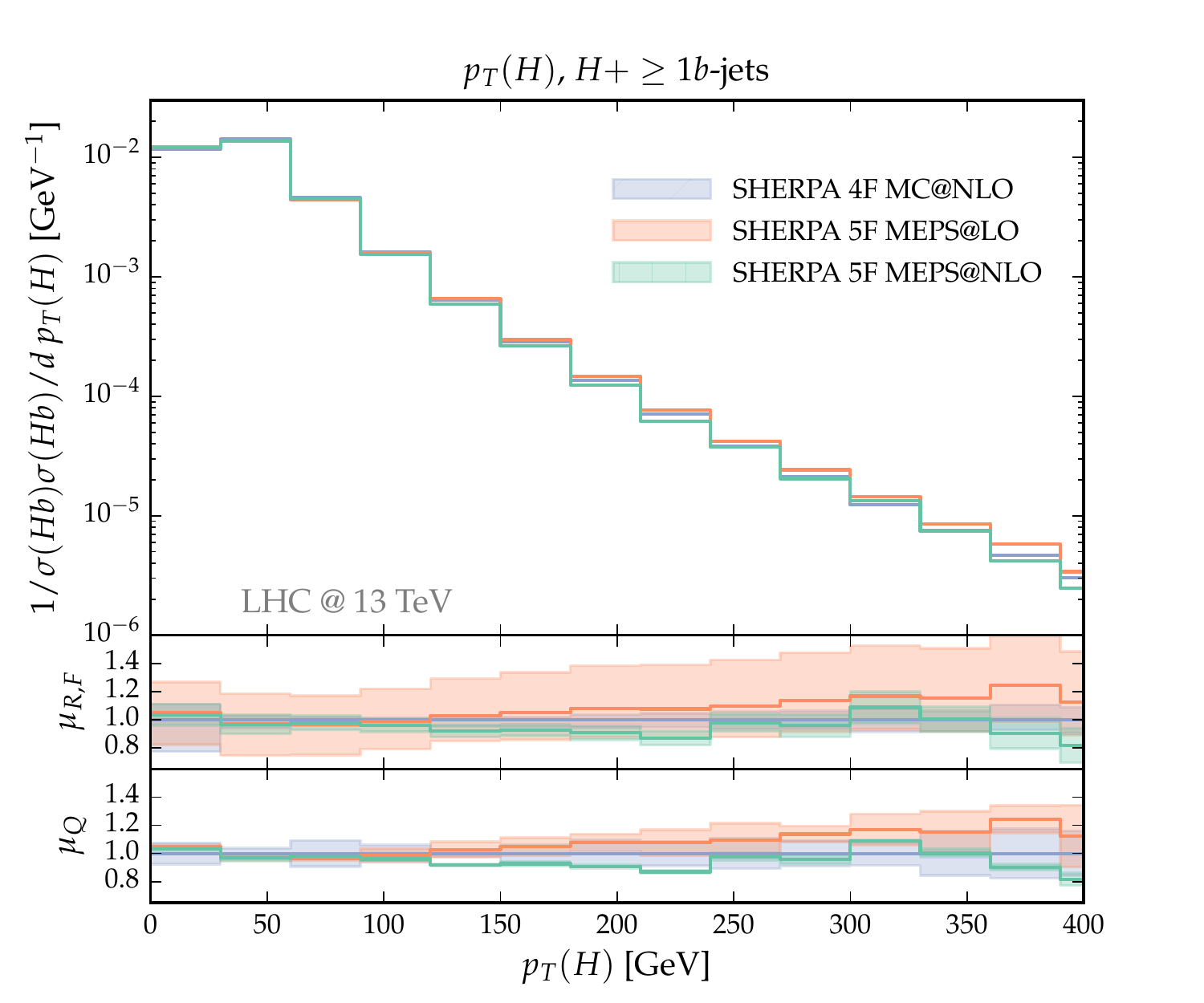}
    \includegraphics[width=0.45\textwidth]{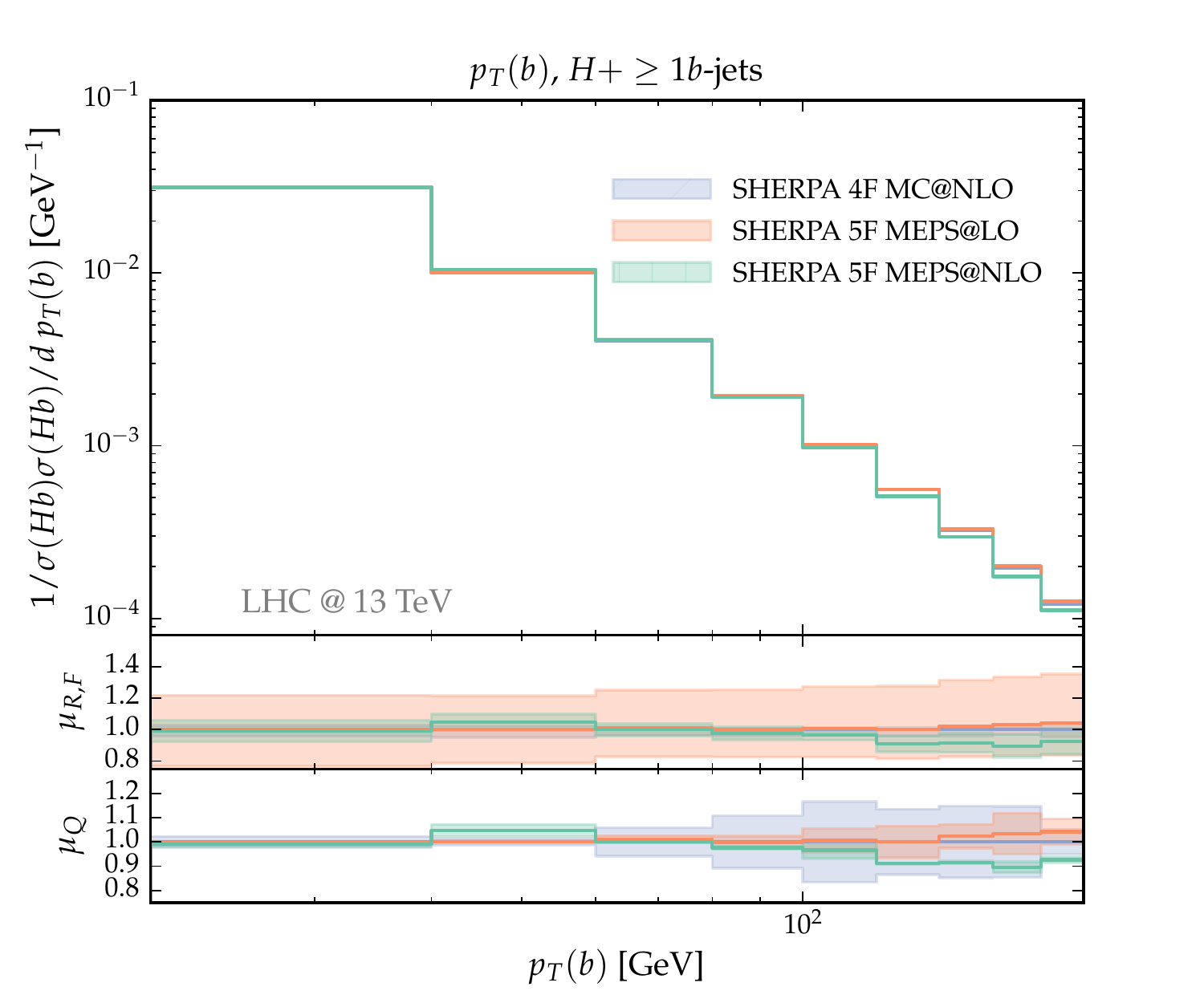}
  }
  \caption{$p_T$ of the $H$ boson and of the leading $b$-jet in the $\geq 1 b$-jet sample.}
  \label{fig:h1b}
\end{figure}

\begin{figure}[htb]
  \centering{
    \includegraphics[width=0.45\textwidth]{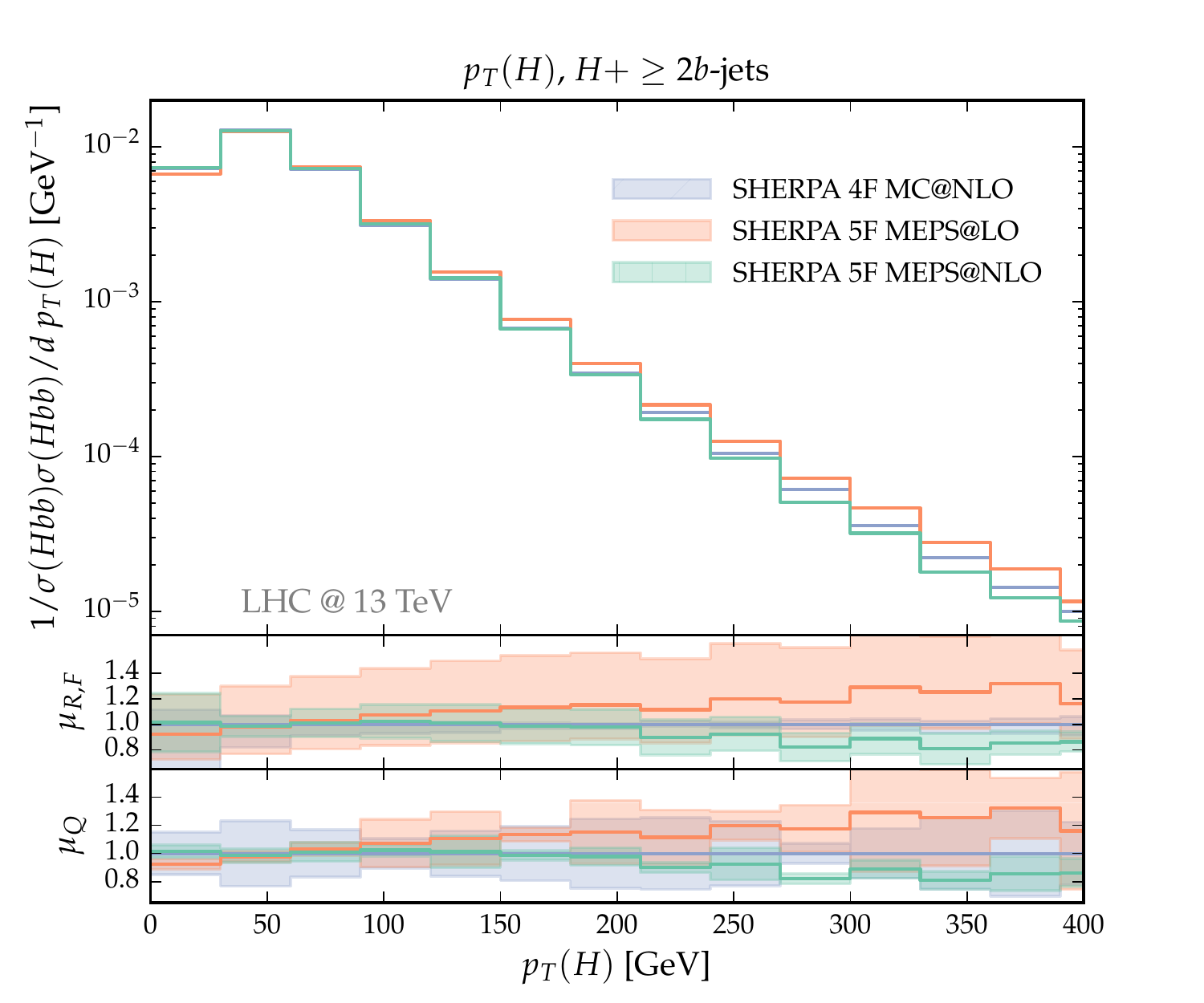}
    \includegraphics[width=0.45\textwidth]{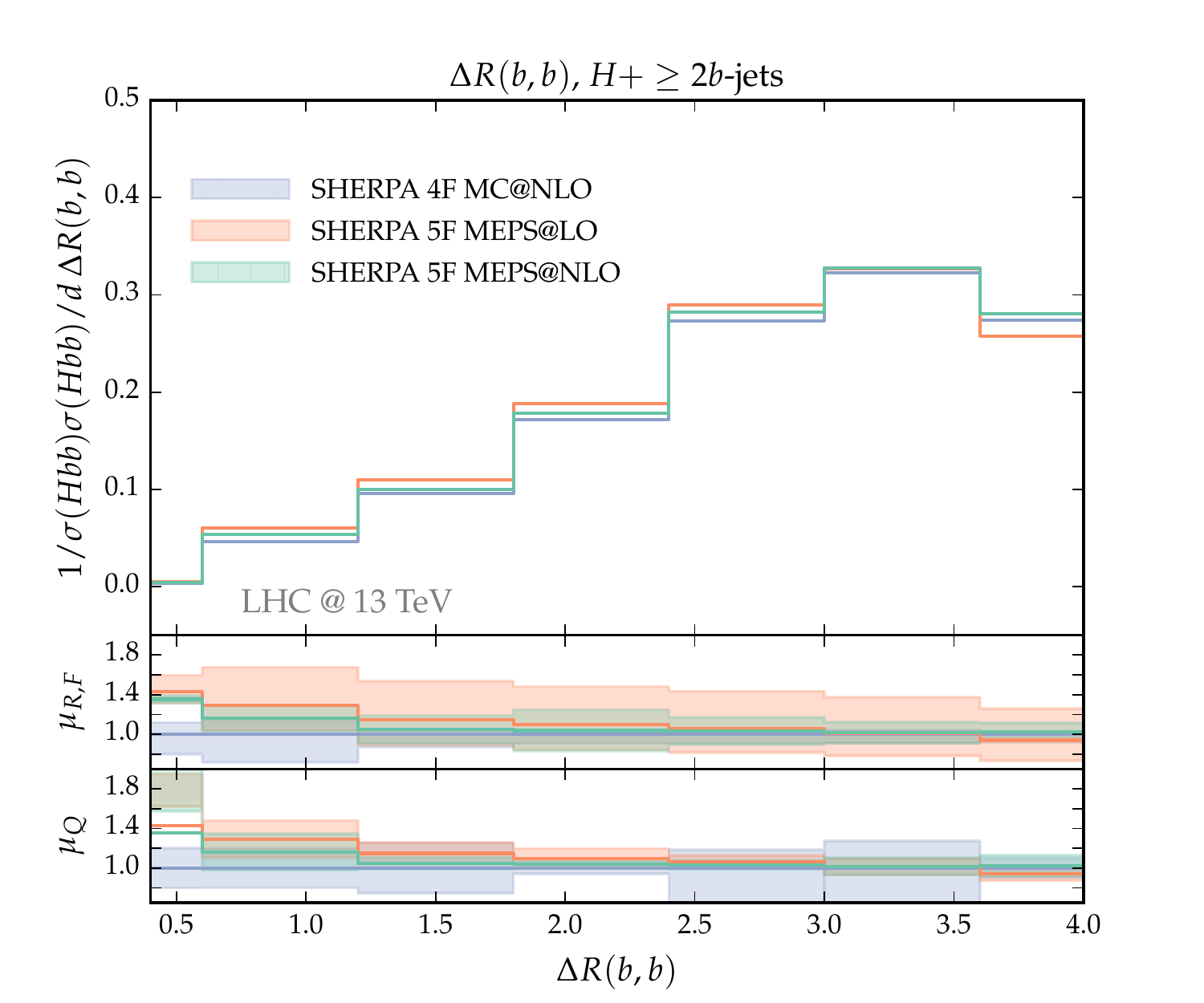}
  }
  \caption{$p_T$ of the $H$ boson and azimuthal distance between the two leading $b$-jets
    in the $\geq 2 b$-jet sample.}
  \label{fig:h2b}
\end{figure}

We can finally extend the results obtained in the previous two sections
to the case of the production of a Higgs boson in association with $b$-jets.
Once again we separate between $\geq 1 b$-jet and $\geq 2 b$-jets samples.
In the following, we exclude normalisation effects as they basically follow
from the previous two sections. Results are shown in Figs.~\ref{fig:h1b}
for samples that exhibit at least one additional $b$-jet, and in Figs.~\ref{fig:h2b}
for samples with at least two $b$-jets tagged in the final state.

The results in the Higgs case seem to lead to the same conclusions
obtained in the $Z$ case, namely that, within theory uncertainties,
the three sample largely agree in shape with the only difference
being due to the normalisation. The only region in which a large ($\sim 40\%$) difference
appears is in the very low $\Delta R(b,b)$ region, where the two $b$-jets can become
collinear in the 5FS. Note that this effect would be largely canceled in samples
that would include fragmentation effects.

\section{Conclusions}
Being able to provide with reliable simulations for LHC processes involving
heavy quarks is necessary in order to precisely determine the background
to many SM processes and BSM searches. This is particularly true for processes
like $Zb\bar{b}$ and $Hb\bar{b}$ in which the choice between the 4F or
the 5FS has historically prove to lead to large discrepancies.

In particular, matching the two schemes, it has been shown that
the difference in the total rate predicted in the two schemes is principally
a consequence of the inclusion of the resummation of initial state logs, and that
mass effects play a very small role.

In addition, with this study, we show that also differences in shapes can be
largely reduce when comparing the two schemes at the same level of accuracy.
We check this claim in the case of $Zb\bar{b}$ production @ 7~TeV, where we have
data to back up our hypothesis. We then extend our results to $Hb\bar{b}$ production @ 13~TeV,
where data are not available. However we can make use of the combined conclusions obtained
for the totally inclusive case and $Zb\bar{b}$, to get some information.
Finally we find that indeed, even in this case, the two schemes do agree in terms of shapes,
with differences being largely within scale uncertainties.

\section*{Acknowledgements}
We want to thank our colleagues from the \Sherpa collaboration 
for fruitful discussions and technical support. We are also thankful
to Stefano Forte, Maria Ubiali and Fabio Maltoni.
We acknowledge financial support from the EU research networks funded by the Research
Executive  Agency (REA) of the European Union under Grant Agreements 
PITN-GA2012-316704 (``HiggsTools'') and PITN-GA-2012-315877 (``MCnetITN''),
by the ERC Advanced Grant MC@NNLO (340983), and from BMBF under contracts
05H12MG5 and 05H15MGCAA,

\end{document}